\documentclass[prd,aps,superscriptaddress,twocolumn,floatfix,nofootinbib]{revtex4}
\pdfoutput=1

\usepackage{amsfonts}
\usepackage{amsmath}
\usepackage{amssymb}
\usepackage{bm}
\usepackage{dcolumn}
\usepackage{graphicx}   
\usepackage[latin1]{inputenc}
\usepackage{latexsym}
\usepackage{rotating}
\usepackage{hyperref}
\usepackage{graphicx}
\usepackage{color}

\newcommand\be{\begin{equation}}
\newcommand\ba{\begin{eqnarray}}
\newcommand\ee{\end{equation}}
\newcommand\ea{\end{eqnarray}}

\begin{document}

\title{Parametric Resonance and Backreaction Effects in Magnetogenesis from Ultralight Dark Matter}

\author{Nirmalya Brahma}
\email{nirmalya.brahma@mail.mcgill.ca}
\affiliation{Department of Physics, McGill University, Montr\'{e}al,
  QC, H3A 2T8, Canada}
\affiliation{Trottier Space Institute, Department of Physics, McGill
University, Montr\'{e}al, QC, H3A 2T8, Canada}
  
\author{Robert Brandenberger}
\email{rhb@physics.mcgill.ca}
\affiliation{Department of Physics, McGill University, Montr\'{e}al,
  QC, H3A 2T8, Canada}
\affiliation{Trottier Space Institute, Department of Physics, McGill
University, Montr\'{e}al, QC, H3A 2T8, Canada}


\begin{abstract}

We take a more detailed look at the recently proposed magnetogenesis mechanism triggered by ultralight dark matter coupled to electromagnetism.  The proposed mechanism made use of a tachyonic resonance channel which leads to the exponential amplification of infrared modes.  Here,  we first investigate a possible narrow band parametric resonance channel which can produce photons at higher frequencies.  Secondly,  we estimate the effects of back-reaction on terminating the resonance. We find that there is indeed a narrow resonance channel. It is characterized by a Floquet exponent which is slightly smaller than the corresponding exponent for the tachyonic resonance.  However, there is a region of parameter space (corresponding to a very small coupling constant) for which the tachyonic resonance is ineffective. In this case, the narrow resonance will dominate, and it will still be sufficiently strong to generate the observed magnetic fields on cosmological scales. Our analytical treatment of the back-reaction effects considered here indicates that a fraction of order one of the initial dark matter density can flow into the gauge fields.  Hence,  our magnetogenesis scenario appears to be robust to back-reaction effects.

\end{abstract}

\maketitle

\section{Introduction} 
\label{sec:intro}

Recently \cite{BJF}, a new mechanism for the generation of magnetic fields on cosmological scales has been proposed \footnote{For an extension to an oscillating scalar dark matter field see \cite{Vahid}.}.  Assuming that the dark matter consists of a pseudoscalar field $\phi$ which is oscillating uniformly on cosmological scales \footnote{For a justification of this assumption - a new derivation of the misalignment mechanism - see \cite{RHB}.} and which is coupled via a axion-like $\phi F \wedge F$ coupling to the field strength tensor $F$ of electromagnetism, it was pointed out that there is a tachyonic resonance which leads to exponential excitation of infrared modes of the electromagnetic vector potential $A_{\mu}$ and hence the generation of magnetic fields.  The instability sets in once the effects of the primordial plasma become inefficient, i.e. at the time of recombination. The Lagrangian of the system (neglecting nonlinear terms in the potential for $\phi$) was taken to be
 \be
 {\cal{L}} \, = \, \frac{1}{2} \partial_{\mu} \phi \partial^{\mu} \phi - \frac{1}{2} m^2 \phi^2 - \frac{1}{4} g_{\phi, \gamma} F \wedge F \, ,
 \ee
 where $g_{\phi \gamma}$ is the constant coupling $\phi$ to $F \wedge F$ and has inverse mass dimensions.  With ultralight dark matter \footnote{See e.g. \cite{Elisa} for reviews.} in mind, a convenient parametrization of the constants is
 \be
 m \, \equiv \, m_{20} 10^{-20} {\rm eV} \, ,
 \ee
 where recent studies provide a bound of $m_{20} > 10$ if $\phi$ makes up all of the dark matter \cite{Dalal},  and
 \be
 g_{\phi \gamma} \, \equiv \, {\tilde{g}}_{\phi \gamma} 10^{-10} {\rm GeV}^{-1} \, .
 \ee
 Observational and experimental bounds on ${\tilde{g}}_{\phi \gamma}$ are given in \cite{Marsh} (see also \cite{site}), and based on these results we expect ${\tilde{g}}_{\phi \gamma} \ll 1$.
 
 Given the above Lagrangian, the mode equation for the amplitude ${\cal{A}}$ of the Fourier mode of the electromagnetic gauge field $A_{\mu}$ becomes
 \be \label{modeeq}
 \bigl( \partial_{\eta}^2 + k^2 \pm k g_{\phi \gamma} a(\eta) {\dot{\phi}} \bigr) {\cal{A}}_{\pm} \, = \, 0 \, ,
 \ee
where the $\pm$ indicate the two different polarization modes of the gauge field.  The function $a(\eta)$ is the cosmological scale factor in terms of conformal time $\eta$, and if we set $a = 1$ today then the comoving momenta correspond to physical momenta today.  Note that this equation assumes that there is no cosmological plasma and hence does not apply before the time of recombination.  After recombination a residual ionization persists \cite{residual}. However, as we show in an appendix, the distance that electrons can travel during the typical time scale of the instability is shorter than the mean free path due to scattering. Hence, for physics which operators on the time scale of the instability the residual ionization can be neglected. On the other hand, the scattering time is somewhat shorter than the time scale of cosmological expansion. Hence, on these longer time scales the inverse cascade mechanism can be used.

Since $\phi$ is oscillating coherently about $\phi = 0$,  i.e.
\be
\phi(t) \, = \, \Phi {\rm{sin}}(mt) \, ,
\ee
where $\Phi$ is the initial amplitude of the oscillations of $\phi$, we see immediately that for modes with wavenumber $k < k_c$ with
 \be \label{kc}
 k_c(\eta) \, \simeq \, g_{\phi \gamma} m \Phi a (\eta) \, . 
 \ee
 there is a tachyonic instability (see also \cite{previous} for earlier studies which made use of this resonance) which leads to an exponential amplification of $A_{\mu}$. In each half period of oscillation it is a different polarization mode which is amplified. In the other half period, the amplitude of the field remain constant, and thus there is net exponential growth of $A_{\mu}$. The rate of exponential growth \footnote{We express the growth as a function of the time variable $t$, i.e. $A_{\mu}(t) \sim e^{\mu_T t}$.} (Floquet index) $\mu_T$ (where the index stands for the tachyonic channel) is
\be \label{Floquet}
\mu_T \, \sim \, g_{\phi \gamma} m \Phi \, .
\ee
 The criterion for being able to neglect the expansion of space is $\mu_T > H$, where $H$ is the Hubble constant at the time of reheating. Also, the scale factor above is the scale factor at that time.
 
 Assuming that the resonance continues until a fraction $F$ of the initial energy in $\phi$ has been converted to photons, the resulting magnetic field on the scale $k_c$ is of the order
 \be
 B(k_c, t_i) \, \sim \, F^{1/2} {\rm{Gauss}} \, ,
 \ee
where $t_i$ is the time when the magnetic field is generated, i.e. shortly after the time of recombination. Since both the phase space of infrared modes and the Floquet exponent (the exponent which describes the exponential increase in the amplitude) are peaked at $k_c$, most of the energy is initially on fluctuations on the scale $k_c$.  The initial magnetic field production process also leads to power on length scales larger than $k_c^{-1}$, but with an amplitude which is Poisson suppressed for a causal generation process \cite{Durrer, Kandu}, i.e.
\be \label{inspec}
B(k,t_i) \, \sim \, \bigl( \frac{k}{k_c} \bigr)^{5/2} B(k_c, t_i) 
\ee
for $k < k_c$. Via an inverse cascade,  the power gradually shifts to lower momenta.  In Minkowski spacetime,  then for the initial spectrum with the power given by (\ref{inspec}), the peak wavelength scales as $t^{2/7}$ and the peak amplitude as $t^{-5/7}$. Hence, at a later time $t$, the strength of the magnetic field at a fixed comoving scale $k$ (e.g. the value of $k$ corresponding to $1 {\rm{Mpc}}$) would be
\be
B(k) \, \sim \,  \bigl( \frac{t_i}{t} \bigr)^{5/7} \bigl( \frac{k}{k_c(t)} \bigr)^{5/2} \, 
= \, \bigl( \frac{t_i}{t} \bigr)^{1/7} \bigl( \frac{k}{k_c} \bigr)^{5/2} B(k_c, t_i)
\ee
where $k_c(t)$ is the peak wavenumber at the time $t$,  and in the final step we have used the scaling of the peak wavenumber mentioned above (see \cite{Kandu} for a review).  In a matter dominated phase of expansion, the wavelength and peak amplitudes scale as a function of a rescaled conformal time instead of as a function of $t$. This rescaled time depends only logarithmically on $t$. Thus, effectively the magnetic field is frozen in as a function of comoving momenta $k$.  In terms of physical coordinates, the strength of $B$ will redshift as $a(t)^{-2}$. It was shown \cite{BJF} that (provided $F$ is not many orders of magnitude smaller than 1) sufficiently strong magnetic fields on a scale of $1 {\rm Mpc}$ can be generated to satisfy the observational lower bound coming from blazar considerations \cite{Vovk}.

There are many open issues concerning this proposed mechanism.  Firstly, as pointed out by one of the authors (NB), there is the possibility of a narrow band parametric resonance channel which could lead to an excitation of $A_k$ modes with $k \simeq k_p \gg k_c$, and that this process would lead to a more rapid conversion of the energy stored in the coherent $\phi$ oscillations. Secondly, in \cite{BJF} no attempt was made to estimate the value of the constant $F$. These are the two questions which we address in this paper.

\section{Possible Narrow Band Parametric Resonance}
 
 The equation (\ref{modeeq}) is that of a harmonic oscillator with an oscillatory component to the mass.  This equation is the famous Mathieu equation, the same equation which also describes the energy transfer from the inflaton field $\phi$ to matter fields $\chi$ at the end of the period of slow roll inflation.  Thus, we can make use of the vast body of work on reheating to apply to our magnetogenesis problem.
 
 In the context of inflationary reheating, it was initially realized \cite{TB, DK} that, even for weak coupling, there is a parametric resonance instability by which modes of $\chi$ can exponentially grow and thus drain the energy from the inflaton.  Later, it was realized that there is often a broad resonance region which dominates \cite{KLS1, STB, KLS2}, and the name ``preheating'' was coined \cite{KLS1} to describe this process. There is the possibility of a tachyonic resonance \cite{tachyonic}.  For reviews of reheating the reader is referred to \cite{ABCM} and \cite{Karouby}.

Returning to our case,  we will, as done in \cite{TB} in the context of reheating, work under the hypothesis that the resonance is fast on the expansion time scale (at the end of the analysis we will check for self-consistency of this assumption). In this case, we can neglect the expansion of space and set conformal time $\eta$ equal to physical time $t$. We can then introduce the dimensionless time variable
\be
z \, \equiv \, m t \, .
\ee
In this case, denoting the derivative with respect to $z$ by a prime, the mode equation (\ref{modeeq}) becomes
\be \label{EoM}
{\cal{A}}_k^{\prime \prime} + [A_k \pm 2 q_k {\rm{cos}}(z)] {\cal{A}}_k \, = \, 0 \, ,
\ee
with
\ba
A_k \, &=& \, \bigl( \frac{k}{am} \bigr)^2 \, \\
q_k \, &=& \, \frac{1}{2} k g_{\phi \gamma} (am)^{-1} \Phi \nonumber
\ea
where $\Phi$ is the amplitude of the oscillation of the field $\phi$.  Assuming that $\phi$ makes up most of the dark matter at the time of recombination we have
\be
m \Phi \, \sim \, T_R^2 \, ,
\ee
where $T_R$ is the temperature at the time of recombination. Th equation (\ref{EoM}) has the form of the Mathieu equation \cite{Mathieu}
\be \label{EoM2}
{\ddot{x}} + \bigl( \delta + \epsilon {\rm{cos}}(t) \bigr) x \, = \, 0 \, 
\ee
with constants
\ba
\delta \, &=& \, \bigl( \frac{k}{am} \bigr)^2 \, \\
\epsilon \, &=& \, k g_{\phi \gamma} (am)^{-1} \Phi \, .
\ea
The region of tachyonic resonance band is given by
\be
\delta \, < \, \epsilon \, ,
\ee
i.e. $k < k_c$.

However, as is well known, there is a narrow parametric resonance band centered at the value of $\delta$ given by
\be
\frac{1}{4} \, = \, \delta \, = \, \bigl( \frac{k}{am} \bigr)^2 \, 
\ee
i.e. $k \equiv k_p = am / 2$, with width $\Delta \delta$
\be
\Delta \delta \, = \, \epsilon \, ,
\ee
i.e. 
\be
\Delta k \, = \, \frac{(am)^2}{2k} \epsilon \, = \, \frac{1}{2} am \Phi g_{\phi \gamma} \, .
\ee
The Floquet exponent $\mu_p$ (where the index indicates we are talking about the parametric resonance channel) is given by \footnote{Note that we are again, as in the case of the tachyonic resonance, expressing the growth as a function of physical time.}
\be
\mu_p \, = \, \frac{1}{2} m \epsilon \, = \, \frac{1}{2} a k g_{\phi \gamma}\Phi \, = \, 
\frac{1}{4} g_{\phi \gamma} m \Phi \, .
\ee
Thus, while the Floquet indices for tachyonic resonance and narrow band parametric resonance are of the same order of magnitude, $\mu_p$ is suppressed compared to $\mu_T$ by a numerical factor, and this leads to the conclusion that the narrow band resonance channel will be less efficient than the broad band one, provided that the phase space of modes undergoing tachyonic resonance is not much smaller than the phase space of modes in the narrow resonance band. This conclusion agrees with the lessons we have learned from inflationary preheating.

Note that if it were not for the different Floquet exponents,  the energy transfer through the tachyonic channel and the narrow band resonance channel could be larger in spite of the narrowness of the resonance band. The reason for this is that the phase space of modes which undergo narrow  resonance is of the order $m^2 \Delta k$ which can be larger than the phase space $k_c^3$ of tachyonic modes for $k_c \ll m$. Concretely, the energy transfer via the tachyonic channel is
\ba
\rho_T \, &\sim& \, k_c k_c^3 e^{2 \mu_T t} \, \sim \, m^4 g_{\phi \gamma}^4 \Phi^4 e^{2 \mu_T t} \nonumber \\
&\sim& \, {\tilde{g}}_{\phi \gamma}^4 e^{2 \mu_T t} 10^{-76} {\rm eV}^4
\ea
while the energy transfer via the narrow resonance channel is
\ba
\rho_{{\rm{narrow}}} \, &\sim& \, k_p k_p^2 \Delta k e^{2 \mu_p t} \, \sim \, m^3 \Delta k e^{2 \mu_p t} \nonumber \\
&\sim& m_{20}^3 {\tilde{g}}_{\phi \gamma} e^{2 \mu_p t} 10^{-80} {\rm eV}^4 \, .
\ea
and thus
\be
\frac{\rho_{{\rm{narrow}}}}{\rho_T} \, \sim \, {\tilde{g}}_{\phi \gamma}^{-3} m_{20}^3 10^{-4} e^{2 [ \mu_p - \mu_T] t} 
\ee
showing that due to the difference in Floquet exponents most of the energy transfer proceeds via the tachyonic channel unless $m_{20} \gg 1$ or ${\tilde{g}}_{\phi \gamma} \ll 1$.

In fact, if $k_c$ is smaller than the Hubble constant at the time of recombination, then the tachyonic resonance becomes ineffective and only the narrow band parametric resonance channel survives.  A rough estimate shows that this will be the case if ${\tilde{g}}_{\phi \gamma} < 10^{-9}$.  The narrow resonance instability may be strong enough to generate magnetic fields on cosmological scales of sufficient amplitude to explain observations, provided $m_{20}$ is not too large. Assuming that a fraction $F$ of the order one of the initial dark matter energy density is transferred to photons, we can estimate the strength to obtain (before redshifting)
\be
B(k_1) \, \sim \, \bigl( \frac{k_1}{k_p} \bigr)^{5/2} 1 {\rm{Gauss}} \, \sim \, m_{20}^{-5/2} 10^{-15} {\rm{Gauss}} \, ,
\ee
where $k_1$ is the value of $k$ corresponding to $1 {\rm{Mpc}}$.

\section{Stability of the Narrow Band Resonance}

 The production of gauge particles will drain energy from the dark matter condensate and thus lead to a decrease in the amplitude of oscillation.  Hence, the amplitude $q_k$ of the oscillation term in the equation of motion (\ref{EoM}) will cease to be constant. This effect does not alter the presence of the tachyonic instability. All that will happen is that the value of $k_c$ slowly decreases. On the other hand, given that the equation (\ref{EoM}) is now no longer a Mathieu equation, and we need to worry about the persistence of the narrow resonance band. From work in the context of inflationary reheating \cite{KLS2} we know that the damping of the oscillations of $\phi$ due to the expansion of space do not destroy the resonance.  Via a rescaling of the matter field $\chi$, the equation can be brought back to the form of a Hill equation \cite{Hill} which admits resonance bands with exponential growth.  In the case of inflationary reheating, matter particle production leads to an additional decrease in the amplitude of $\phi$, and this is a similar effect to the one we are considering now.  We can call this effect a {\it back-reaction effect}.  From studies of inflationary reheating (see e.g. \cite{ABCM}) we know that back-reaction eventually shuts off the resonance. In this section we want to estimate when back-reaction will terminate the resonance.
 
 There are a number of back-reaction channels to consider. Firstly, on energetic grounds the production of $A_{\mu}$ particles will lead to a decrease in the amplitude of $\phi$. Secondly,  the production of $A_{\mu}$ quanta will lead to the generation of inhomogeneities in $\phi$ which destroy the coherence of the $\phi$ condensate which is crucial for the resonance. We will consider these effects in turn.
 
We first study the effects of a slow decrease in the amplitude of $\phi$. We write $\epsilon = \epsilon(t)$. When Taylor expanding about $t  = 0$ we can write
\be
\epsilon(t) \, = \, \epsilon_0 ( 1 - \alpha t)    \, .
\ee
To see how this change effects the resonance, we will adapt the analytical approach in \cite{Rand} \footnote{See also the physics textbook discussions in \cite{LL, Arnold}.}(which is called the ``two time method'' in that reference).

The starting point is Eq. (\ref{EoM2})
\be \label{EoM3}
{\ddot{x}} + \bigl( \delta + \epsilon(t) {\rm{cos}}(t) \bigr) x \, = \, 0 \, ,
\ee
where we have added the time dependence of $\epsilon$. We define a second time variable (the ``slow time'') $\eta \, \equiv \, \epsilon t$  and make the ansatz
\be
x \, = \, x_0(t, \eta) + \epsilon(t) x_1(t) \, ,
\ee
where $x_0$ is a solution of the equation (\ref{EoM3}) for $\epsilon = 0$ with coefficients which depend on the slow time
\be
x_0(t, \eta) \, = \,  A(\eta) {\rm{cos}}(\sqrt{\delta} t) + B(\eta) {\rm{sin}}(\sqrt{\delta} t) \, .
\ee
The time dependence of the coefficients $A$ and $B$ is determined such that the correction term $x_1$ remains small.  Taking the second derivative of $x_0$ yields
\ba
\frac{d^2x_0}{dt^2} \, &=& \, \bigl( {\dot{\epsilon}} \frac{\partial A}{\partial \eta} + \epsilon^2 \frac{\partial^2 A}{\partial^2 \eta} \bigr) {\rm{cos}} (\sqrt{\delta} t) - 2 \epsilon \frac{\partial A}{\partial \eta} \sqrt{\delta} {\rm{sin}}(\sqrt{\delta} t) \nonumber \\
&-& \, \delta A {\rm{cos}}(\sqrt{\delta} t)  \\
&+& \, \bigl( {\dot{\epsilon}} \frac{\partial B}{\partial \eta} + \epsilon^2 \frac{\partial^2 B}{\partial^2 \eta} \bigr) {\rm{sin}} (\sqrt{\delta} t) + 2 \epsilon \frac{\partial B}{\partial \eta} \sqrt{\delta} {\rm{cos}}(\sqrt{\delta} t) \nonumber \\
&-& \, \delta B {\rm{sin}}(\sqrt{\delta} t) \nonumber
\ea
Neglecting terms of order $\epsilon^2$ we find
\ba
\frac{d^2x_0}{dt^2} - \delta x_0 \, &=& \,  {\dot{\epsilon}} \frac{\partial A}{\partial \eta}  {\rm{cos}} (\sqrt{\delta} t) - 2 \epsilon \frac{\partial A}{\partial \eta} \sqrt{\delta} {\rm{sin}}(\sqrt{\delta} t) \nonumber \\
 &+& \,  {\dot{\epsilon}} \frac{\partial B}{\partial \eta} {\rm{sin}} (\sqrt{\delta} t) + 2 \epsilon \frac{\partial B}{\partial \eta} \sqrt{\delta} {\rm{cos}}(\sqrt{\delta} t) \, .
 \ea
 These terms are all of the order $\epsilon$. In order for the full equation to be satisfied to order $\epsilon$, the equation for $x_1$ becomes
 \ba
 {\ddot{x_1}} &+& \delta x_1 \, = - \epsilon(t) {\rm{cos}}(t) x_0 \\
 &-&  {\dot{\epsilon}} \frac{\partial A}{\partial \eta}  {\rm{cos}} (\sqrt{\delta} t) - 2 \epsilon \frac{\partial A}{\partial \eta} \sqrt{\delta} {\rm{sin}}(\sqrt{\delta} t) \nonumber \\
 &-&  {\dot{\epsilon}} \frac{\partial B}{\partial \eta} {\rm{sin}} (\sqrt{\delta} t) + 2 \epsilon \frac{\partial B}{\partial \eta} \sqrt{\delta} {\rm{cos}}(\sqrt{\delta} t) \, .\nonumber
 \ea
 Inserting the ansatz for $x_0$ in the first term on the right hand side of this equation, using identities for products of trigonometric functions, focusing on the central value of the resonance band $\delta = 1/4$ and dropping terms which oscillate faster than $cos(t/2 + \alpha)$ ($\alpha$ being a phase) we obtain
 \ba
  {\ddot{x_1}} + \delta x_1 \, &=& \, \bigl( \frac{\partial A}{\partial \eta} + \frac{1}{2}B + \frac{{\dot{\epsilon}}}{\epsilon} \frac{\partial B}{\partial \eta}\bigr) {\rm{sin}}(t/2) \nonumber \\
  &-& \bigl( \frac{\partial B}{\partial \eta} + \frac{1}{2}A + \frac{{\dot{\epsilon}}}{\epsilon} \frac{\partial A}{\partial \eta}\bigr){\rm{cos}}(t/2) \, . \nonumber
 \ea
 The idea of the method is the realization that in order for $x_1$ to remain small, the coefficients of the two oscillatory terms must cancel. Hence we have (with a prime denoting the derivative with respect to $\eta$)
 \ba \label{coeff1}
 A^{\prime} + \frac{1}{2}B + \frac{{\dot{\epsilon}}}{\epsilon} B^{\prime} \, = \, 0 \nonumber \\
 B^{\prime} + \frac{1}{2}A + \frac{{\dot{\epsilon}}}{\epsilon} A^{\prime} \, = \, 0 \, .
 \ea
 In the case of constant $\epsilon$ we can combine the two equations to obtain
 \be \label{coeff2}
 A^{\prime \prime} \, = \, \frac{1}{4} A 
 \ee
 which shows that $x_0(t)$ grows exponentially with Floquet index $\epsilon / 2$. When $\epsilon$ is not constant we can make the ansatz that the coefficients $A$ and $B$ grow exponentially in $\eta$ with a coefficient $\lambda$. The resulting equation for $\lambda$ becomes (neglecting terms of order ${\dot{\epsilon}}^2$)
 \be
 \lambda^2 - \frac{1}{2} \frac{{\dot{\epsilon}}}{\epsilon} \lambda - \frac{1}{4} \, = \, 0 \, . 
 \ee
 This shows that a small time dependence of $\epsilon$ does not destroy the resonance, but only leads to a small decrease in the exponent.
 
 In \cite{Rand}, the above method was used to (in the case of constant $\epsilon$) determine the width of the resonance band. We set $\delta = 1/4 + \delta_1$, where $\delta_1$ is of the order $\epsilon$. This leads to correction terms in the coefficient equations (\ref{coeff1})and (\ref{coeff2}). We then determine the maximal value of $\delta_1$ for which exponential solutions of (\ref{coeff2}) persisit. We can extend this analysis to the case of time-dependent $\epsilon$ (as done above for $\delta = 1/4$) and we find that the width of the instability band is only modified by a quantity of the order of ${\dot{\epsilon}} / \epsilon$.
 
 However, a warning is warranted concerning this analysis: it is an approximate study - perturbative in $\epsilon$ and neglecting more rapidly oscillating term. Even the condition for resonance is not rigorous. In the case of constant $\epsilon$, there is a rigiorous mathematical theory (Floquet theory \cite{Floquet}) which supports the conclusions. For slowly decreasing $\epsilon$ we do not have such mathematical support. Hence,  it is useful to turn to a numerical study.

\section{Backreaction Effects and the Termination of the Resonance} \label{sec:Bfield}

A key question is at what point the resonance will terminate.  Obviously, there is a basic energetic requirement: the resonance cannot drain more energy from the condensate than it had initially. Let us perform an order of magnitude analysis of the duration of the resonant process based on this criterion.  Focusing on the tachyonic resonance channel, the energy density in the produced ${\cal{A}}_{\mu}$ fields is
\be
\rho_A \, \sim \, \int_{|k| < k_c} d^3k k^2 {\cal{A}}_{\mu, k} {\cal{A}}^{\mu}_k \, ,
\ee
where the subscript $k$ stands for the Fourier mode. Using vacuum initial conditions for ${\cal{A}}_{\mu}$, namely $|{\cal{A}}_{\mu}| = 1 / \sqrt{2k}$, this gives (making use of the fact that most of the energy goes into modes with $k \sim k_c$
\be 
\rho_A(t) \, \sim \, k_c^4 e^{2 \mu_T t} \, ,
\ee
where we set the clock such that $t = 0$ is the onset of the resonance.  Making use of the expression (\ref{kc}) for $k_c$, and of the assumption that at $t = 0$ the energy density in $\phi$ is the total dark matter density at recombination $\tau$ of the resonance
\be \label{bound1}
e^{2 \mu_T \tau} \, \sim \, {\tilde{g}}_{\phi \gamma}^{-4} 10^{76} \, 
\ee
or
\be
\tau \, \sim \, \frac{1}{2} \mu_T^{-1} [ 4 {\rm{log}} {\tilde{g}}_{\phi \gamma}^{-1} + 76 {\rm{log}} 10 ] \, .
\ee
The condition that the expansion of space is negligible is $\tau < H^{-1}$ (where $H$ is the expansion rate at recombination).  This yields a lower bound on ${\tilde{g}}_{\phi \gamma}$ which is of the order
\be
{\tilde{g}}_{\phi \gamma} \, > \, 10^{-4} \, ,
\ee
where we have made use of the fact that at recombination $a \sim 10^{-3}$. Note that there is still resonance when this bound is not satisfied, but that then the amplitude of the gauge field is also damped by the expansion (superimposed on the tachyonic growth). The intuition for this is the same as in the case of inflationary preheating (see e.g. the review in \cite{ABCM}). We conclude that the energy criterion does not prevent a fraction $F = {\cal{O}}(1)$ of the initial energy density of dark matter to be transferred to gauge fields.

Let us now study constraints which come from demanding that back-reaction does not destroy the coherence of the condensate. To do that, we must consider the Klein-Gordon equation for $\phi$
\be \label{KG}
{\ddot{\phi}} + 3 H {\dot{\phi}} - \nabla^2 \phi + m^2 \phi \, = \, g_{\phi \gamma} F \wedge F \, .
\ee
We are interested in studying both the corrections to the homogeneous part of $\phi$, and the growth of inhomogeneities in $\phi$. For this purpose, we expand the fields in Fourier modes and work in the $A_0 = 0$ gauge
\ba
\phi(x, t) \, &=& \, \phi_0(t) + \delta \phi(x, t) \\
\delta \phi(x,t) \, &=& \, \int d^3k e^{ikx} a_k \delta \phi(k, t) \nonumber \\
A_i(x, t) \, &=& \, \int d^3k e^{ikx} {\cal{A}}_i(k,t) b_{i, k}\, \nonumber
\ea
where by definition $\delta \phi$ does not contain any homogeneous mode. In the above, $a_k$ and $b_{i, k}$ are canonically normalized mode creation operators. When evaluating correlation functions, we will be quantum averaging and making use of the canonical commutation relations.

We can extract the correction to the zero mode of $\phi$ by spatially averaging (\ref{KG}):
\be \label{KGhom}
{\ddot{\phi}}_0 + 3 H {\dot{\phi}}_0  + m^2 \phi_0 \,  = \,
- g_{\phi \gamma} \int d^3k k_j \epsilon^{0jab} {\dot{{\cal{A}}}}_a(k) {\cal{A}}_b(k) \, ,
\ee
where $\epsilon^{ijab}$ is the totally antisymmetric tensor, and the integral runs over all $k$ modes which undergo resonance.  From (\ref{KGhom}) we infer that the order of magnitude of the correction to the homogeneous mode of $\phi$ is given by $m^{-2}$ times the right hand side of (\ref{KGhom}).  Demanding that the correction term remains smaller than $\Phi$ leads to (under the same assumptions as used above) the following upper bound on $\tau$
\be \label{bound2}
 e^{2 \mu_T \tau} \, \sim \, m_{20} {\tilde{g}}_{\phi \gamma}^{-5} 10^{84} \, ,
\ee
which is a weaker bound than the previous one.

Finally, let us consider the energy density in the Fourier modes of $\phi$ which are excited. To obtain the Fourier mode $\delta \phi(k)$ we insert the Fourier expansion of the fields into the of motion (\ref{KG}) and take the quantum expectation value over the gauge field modes, making use of the canonical commutation relations for the $b_{i, k}$.  In this way,  we obtain the estimate
\be \label{relation}
k^2 \delta \phi(k) a_k \, \sim - \, g_{\phi \gamma}  k_j {\dot{{\cal{A}}}}_a(k) {\cal{A}}_b(- k) \epsilon^{0jab} \, ,
\ee
and the expression for the resulting energy density is
\be \label{denspert}
\delta \rho(x) \, = \, \int d^3k d^3k' k k' \delta \phi(k) \delta \phi(k') a_k a_{k'} e^{- ix(k - k')} \, .
\ee
 In order to estimating the total energy density in fluctuations we insert (\ref{relation}) into (\ref{denspert}) and  integrate the contribution of all Fourier modes in the tachyonic resonance band. This yields
\be
\rho_{\delta \phi}(t) \, \sim \, k_c^6 g_{\phi \gamma}^2 e^{4 \mu_T t}
\ee
which yields a bound very similar to the previous one
\be
 e^{2 \mu_T \tau} \, \sim \, {\tilde{g}}_{\phi \gamma}^{-4} 10^{85} \, .
\ee
We thus conclude that based on these analytical considerations the development of $\phi$ fluctuations does not terminate the resonance before a fraction $F = {\cal{O}}(1)$ of the energy density has transformed.

\section{Numerical Study}

In light of the many approximations made in the previous analysis, it is important to study the dynamics numerically.  We have made an initial numerical study in which we study the homogeneous equation of motion for $\phi$ coupled to the gauge field equation  Making use of the dimensionless time variable $z = mt$ introduced earlier, the equations to be solved are
\be \label{modeeq2}
\bigl[ \partial_z^2 + \kappa^2 \pm g_{\phi \gamma} \kappa \partial_z \phi \bigr] {\cal{A}}_{\pm}(\kappa) \, = \, 0 \, ,
\ee
with $\kappa = k/(am)$, and
\ba \label{KGhom2}
\bigl[ \partial_z^2 + 1 \bigr] \phi_c(z) \, &=& \, \frac{g_{\phi \gamma}}{a} m^3 
\int d^3 \kappa^\prime \frac{\kappa^\prime}{(2\pi)^3} \\
& & \, \bigl({\cal{A}}_+^*(\kappa^\prime) \partial_z{\cal{A}}_+(\kappa^\prime) - {\cal{A}}_-^*(\kappa^\prime) \partial_z{\cal{A}}_-(\kappa^\prime) \, . \nonumber
\ea

Instead of a full numerical simulation, we focus on two fixed $k$ modes, one ($k = k_T$) corresponding to the tachyonic resonance band, the other ($k = k_p$) to the center of the narrow resonance band.   We evolve (\ref{modeeq2}) for the two modes together with (\ref{KGhom2}), with the integral on the right hand side of (\ref{KGhom2}) replaced by the sum of the contribution $\kappa_T^4 {\cal{A}}(\kappa_T)^* \partial_z {\cal{A}}(\kappa_T)$ of the  tachyonic modes with wavenumbers of the order of $k_T$,  and the contribution of modes with wavenumber in the narrow resonance band (i.e.  $\kappa_p^2 (\Delta \kappa) {\cal{A}}(\kappa_p)^* \partial_z {\cal{A}}(\kappa_p)$).

Figs. 1 and 2 display the evolution in the case of a large coupling constant ${\tilde{g}}_{\phi \gamma} = 5 \times 10^2$ and a mass consistent with the current lower bound, namely $m_{20} = 10^2$.  Figure 3 is for the case of a smaller coupling constant, namely ${\tilde{g}}_{\phi \gamma} = 1$. In each of these plots,  the horizontal axis is the time variable $z$. The top panel shows the growth of the amplitude of ${\cal{A}}_k$ (whose values are mentioned below), the second panel depicts the amplitude the energy density in $\phi$, in the gauge field modes undergoing tachyonic resonance, and in those undergoing narrow resonance growth, and the third panel the time dependence of the effective square mass in the mode equation (which when negative indicates the possibility of tachyonic growth and which oscillate with opposite phases for the two polarization modes - see the brown versus green curves). 

In the top panel, the dotted and dashed straight black lines indicate the growth expected from our analytical analysis for tachyonic resonance (using the maximal value of the Floquet exponent for tachyonic resonance, and the value for the narrow band resonance - which is one quarter of the former -, respectively). The cyan solid and dashed curves give the value of ${\cal{A}}_k$ observed for the value of $k_T$ in the tachyonic band for the two different polarization modes,  and the purple solid and dashed curves are the corresponding ones for the narrow resonance band.  

In the second panel, the brown curve gives the energy density in $\phi$, the cyan curve the energy density in the tachyonic modes, and the purple curve the corresponding energy in the narrow resonance modes.

\begin{figure}
  	\includegraphics[width=8cm]{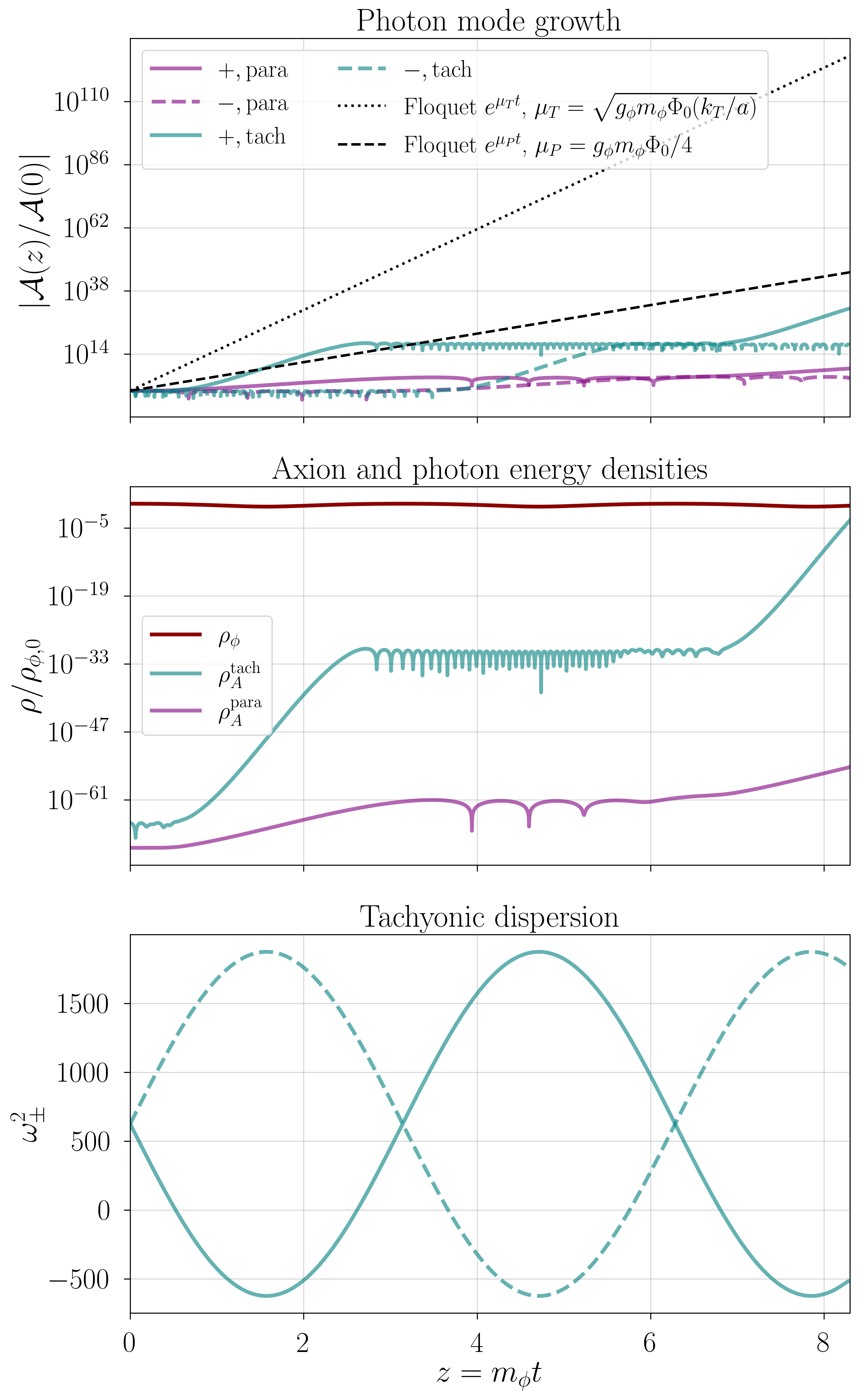}
	\caption{Evolution of the gauge field mode $A_k$ and the background scalar field in the case of large coupling (${\tilde{g}}_{\phi \gamma} = 5 \times 10^2$ - see text for description of the curves), and for $k_T = 0.5k_c$.}
	\label{fig1}
\end{figure}

\begin{figure}
  	\includegraphics[width=8cm]{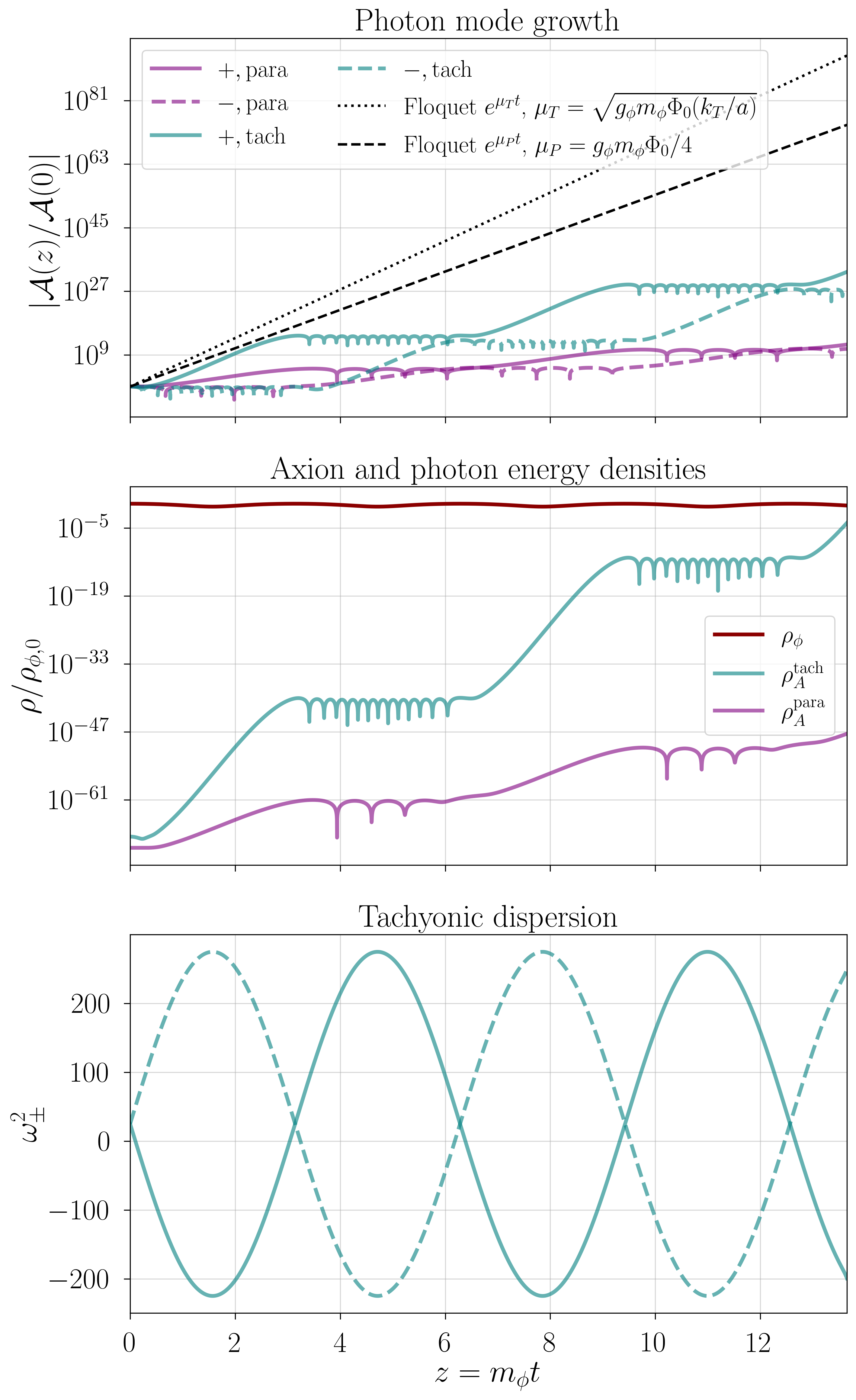}
	\caption{Evolution of the gauge field mode $A_k$ and the background scalar field in the case of large coupling (${\tilde{g}}_{\phi \gamma} = 5 \times 10^2$ - see text for description of the curves), and for $k_T = 0.1k_c$.}
	\label{fig1}
\end{figure}

From Figure 1 we see that for large couplings, the tachyonic resonance dominates,  as is expected from our analytical analysis.   The value of $k_T$ chosen is $k_T = \frac{1}{2} k_c$.  Tachyonic resonance proceeds in bursts, as expected since the growth proceeds only in half of the oscillation period of $\phi$. The slope of the exponential growth agrees with the analytical estimate. However, after a couple of oscillations our code for the evolution of this $k$ mode breaks down.   On the other hand, the tachyonic resonance continues for modes with smaller values of $k$, as demonstrated in Figure 2 where the evolution of the mode with $k_T = 0.1 k_c$ is shown.  The results indicate that there is no reason to believe that the tachyonic resonance terminates before the fraction of energy of $\phi$ converted into gauge fields is of the order $1$,  in agreement with the conclusions from  our analytical analysis.

\begin{figure}
  	\includegraphics[width=8cm]{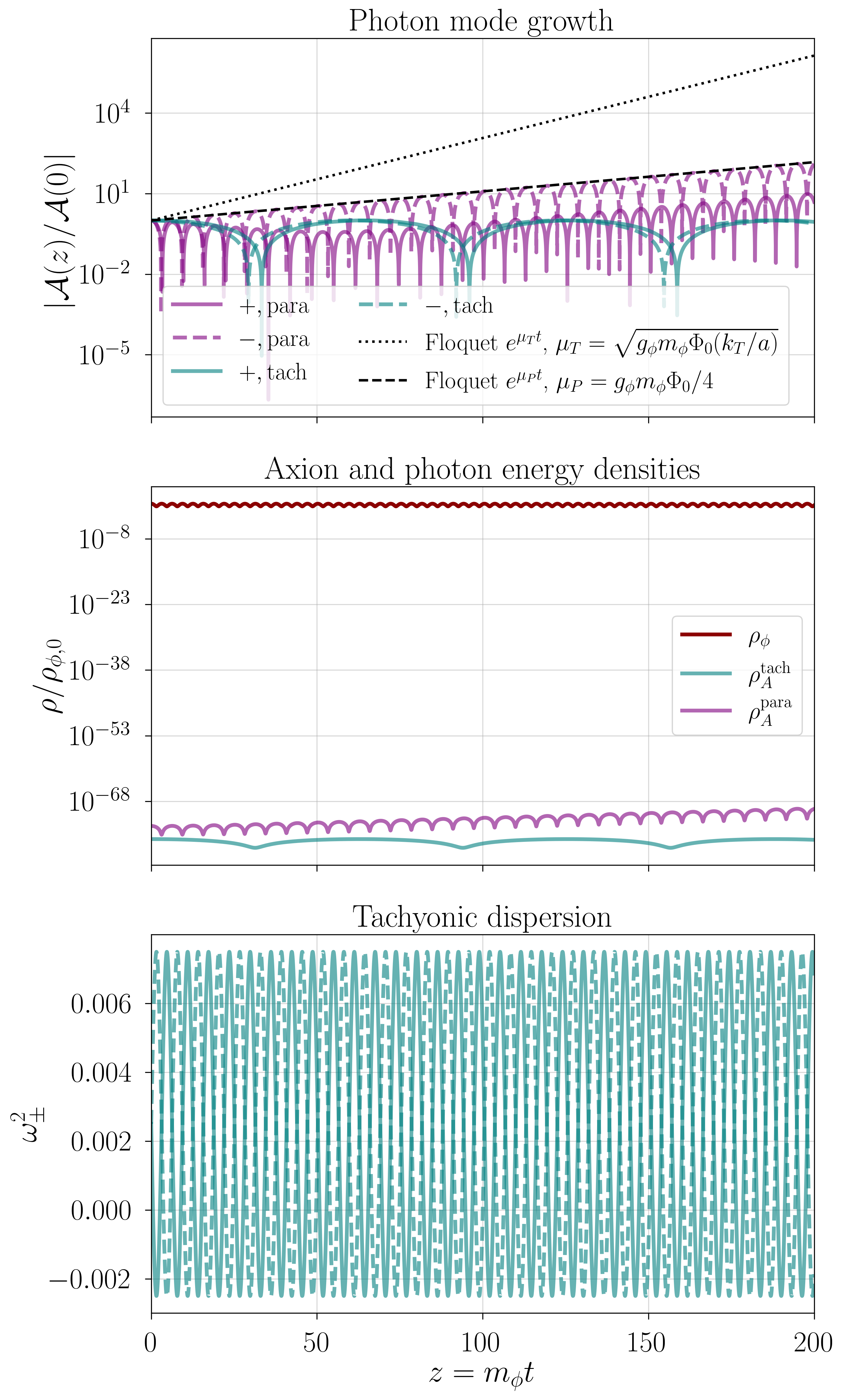}
	\caption{Evolution of the gauge field mode $A_k$ and the background scalar field in the case of small coupling (${\tilde{g}}_{\phi \gamma} = 1$ - see text for description of the curves), and for $k_T = 0.5k_c$.}
	\label{fig1}
\end{figure}

In the case of small coupling constant ${\tilde{g}}_{\phi \gamma} = 1$, the results (see Figure 3) show that the tachyonic resonance is ineffective and the narrow resonance dominates.   The dotted line corresponds to the increase with the Floquet exponent of tachyonic resonance, the dashed line corresponds to the growth with the Floquet exponent of the narrow parametric resonance.  We see an excellent agreement between the numerical growth of ${\cal{A}}_k$ and what is predicted from narrow band parametric resonance.

A criterion for tachyonic resonance to be effective is that the amplitude of ${\cal{A}}_k$ during one burst of particle production is significantly larger than one.  If this is not the case, we cannot neglect the non-trivial matching of the oscillatory evolution of ${\cal{A}}_k$ while $\omega_k^2 > 0$ to the tachyonic growth when $\omega_k^2 < 0$.  The criterium for efficiency of tachyonic resonance is therefore
\be
\frac{1}{4} p \mu_{k} \, \gg \, 1 \, ,
\ee
where $p$ is the period, and the value of $\mu$ is evaluated at $k = k_c$.  This yields the condition
\be \label{cutoff}
{\tilde{g}}_{\phi \gamma} m_{20}^{-1} \, \gg \, 1 \, ,
\ee
which is satisfied in the case studied in Figs. 1 and 2, but not in the case studied in Fig. 3.

\section{Discussion and Conclusions}

 We have studied various aspects of the recently proposed late time magnetogenesis scenario in which a coherently oscillating ultralight dark matter field generates magnetic fields on cosmological scales via a parametric instability.  In previous work \cite{BJF} only the tachyonic channel was studied. Here we have shown that there is an additional narrow resonance band channel by which shorter wavelength magnetic fields could be produced.  The phase space of modes which undergoes the narrow resonance instability is comparable to (and for many parameter values larger than) the phase space which undergo tachyonic resonance, but the Floquet exponent for the narrow band resonance is lower (but only by a numerical factor) than the Floquet exponent for the tachyonic resonance.  Hence,  provided that the tachyonic resonance operates, it will dominate. We have seen that for small values of the coupling constant ${\tilde{g}}_{\phi \gamma}$ the tachyonic resonance shuts off because the increase in the amplitude of the gauge field during each burst of particle production is too small.  In this case only the narrow resonance instability band survives. We have derived a criterion for the cutoff of the tachyonic resonance (see (\ref{cutoff}) and confirmed the result with initial numerical work.
 
We have studied the stability of the narrow resonance regime to back-reaction effects.  Naive analytical approximations indicate that the resonance persists, although this result should be confirmed with a full numerical analysis.
 
 A key question not addressed in earlier work \cite{BJF, Vahid} is the issue of  when back-reaction effects will shut off the resonance, and what the fraction $F$ of the initial dark matter energy is when the resonance shuts off. Our analytical approximations and numerical work indicate that, provided that the tachyonic resonance channel is effective, then a fraction $F \sim 1$ of the energy is converted. This implies that the magnetogenesis mechanism proposed in \cite{BJF} is indeed quite effective at producing cosmological magnetic fields of the required magnitude. On the other hand, if only the narrow resonance channel is open, then a full numerical study is required to determine the value of $F$ (this study is currently in progress).
 
 Note that similar equations to the ones we study here were investigated in detail in the context of gauge preheating after axion inflation \cite{Giblin}.  In that case, the field $\phi$ was a pseudoscalar inflaton,  and the gauge field was the same $U(1)$ electromagnetic field considered here.  Note that the authors of \cite{Giblin} also considered the application of their mechanism to magnetogenesis \cite{Giblin2}.
 
\section*{Acknowledgement}

\noindent 
The research at McGill is supported in part by funds from NSERC and from the Canada Research Chair program.   NB was supported in part by a Doctoral Research Scholarship from the Fonds de Recherche du Qu\'ebec -- Nature et Technologies and by the Canada First Research Excellence Fund through the Arthur B. McDonald Canadian Astroparticle Physics Research Institute. 

\section*{Appendix}

Residual ionization can hinder magnetogenesis, which has been appreciated in the literature (see e.g. \cite{Bassett, Taiwan}).  Here we justify why our study of the resonant production of the magnetic field the residual ionization can be neglected.  

Note that our dynamics yields oscillating electromagnetic fields.  We are not discussing the propagation of the magnetic field into the voids (as process considered in the recent paper \cite{Axel}), but the uniform generation of a magnetic field everywhere in space, and in particular in the voids.  Hence, what is relevant is not the DC conductivity, but the AC conductivity.  

In the case of propagation of the magnetic field into the bulk, the condition for being able to neglect the conductivity in the equations of motion is
\be
L \, \ll \, l_{\rm mfp} \, ,
\ee
where $l_{\rm mfp}$ is the mean free path for Coulomb scattering of electrons - the dominant process which needs to be considered \cite{Axel}, and $L$ is the size of the void.  As shown in \cite{Axel}, this condition is violated at all times, and hence to study the propagation of the magnetic field into the bulk the equations for a collisional plasma (which contain the DC conductivity) need to be used, and one finds that at no time between recombination and the present the propagation into the void is effective \cite{Axel}.

In our case,  we are not discussing the propagation of a magnetic field into the void, but the generation inside the void via an instability with a particular time scale.  Thus, the condition to be able to neglect the conductivity (which would have to be the AC conductivity if it were important) is that electrons during the time scale of the instability should not be able to move a greater distance than $l_{\rm mfp}$.  This condition reads
\be \label{crit}
\mu^{-1} \frac{v_e}{c} \, \ll l_{\rm mfp} \, ,
\ee
where $v_e$ is the root mean square value of the electron velocity, and $\mu$ is the Floquet exponent discussed in Sections I and II of the paper.   The mean free path of electrons is discussed in detail in \cite{Axel}.  From the bottom panel of Fig. 5 in that paper we can read off that at a redshift of $z = 100$
\be
l_{\rm mfp} \, \sim \, 10^{-9} H^{-1} \, ,
\ee
where $H$ is the Hubble expansion rate (which is given via the Friedmann equation by the temperature and Newton's gravitational constant).  Making use of the expression \ref{Floquet} for $\mu$ and taking the value of $m \Phi$ to be given by demanding that $\phi$ is the dominant component of dark matter, yields the condition
\be \label{cond1}
{\tilde{g}}_{\phi \gamma} \, > \, 10^{-4} \, 
\ee
for (\ref{crit}) to be satisfied, and hence to justify neglecting the residual conductivity. Here, we have used the fact that $v_e / c \leq 10^{-4}$ after recombination.

In the case of narrow band resonance, there is a second time scale to consider, namely the oscillation time of $\phi$ which is $m^{-1}$. The condition
\be
m^{-1} \frac{v_e}{c} \, \ll l_{\rm mfp} 
\ee
yields the criterion
\be \label{cond2}
m_{20} \, > \, 10^{-3} 
\ee
which needs to be satisfied (in addition to (\ref{cond1}) in order to justify neglecting plasma terms in the equations of motion.  Note that (\ref{cond2}) is satisfied for all observationally viable axion-like dark matter fields,  and only (\ref{cond1}) remains as a non-trivial constraint.

For completeness we mention an additional consideration. Photons interact with free electrons via Thomson scattering. The scattering time $t_T$ is given by
\be
t_T \, = \, \frac{1}{n_e \sigma_T} \, ,
\ee
where $\sigma_T$ is the Thomson cross section and $n_e$ is the number density of free electrons, which in turn is given by $z_e n_b$, where $n_b$ is the baryon number density and $z_e$ is the residual abundance. Inserting the values of $n_b$ (at the time of recombination) and $\sigma_T$, and using a value $z_e \sim 10^{-4}$ for the residual ionization \cite{residual} we find that right after recombination
\be
t_T \, \sim \, 10^{37} {\rm{GeV}}^{-1} \, .
\ee
On the other hand, the time scale of narrow parametric resonance is given by
\be
k_p^{-1} \, \sim \, (a m)^{-1} \, ,
\ee
where $a$ is the value of the scale factor at recombination (taking $a$ to be normalized to the value $a = 1$ today).  For ultralight dark matter with a mass $m = m_{20} 10^{-20} {\rm{eV}}$ we get
\be
k_p^{-1} \, \sim \, a^{-1} m_{20}^{-1} 10^{29} {\rm{GeV}}^{-1} \, .
\ee
Hence, the condition under which it is self-consistent to neglect the residual ionization on the time scale of our instability is
\be
m_{20} \, > \, 10^{-5} \, ,
\ee
and thus satisfied for all ultralight dark matter fields consistent with the observational constraints.

Note that the scattering time is smaller than the Hubble expansion time. (which is of the order $10^{38} {\rm{GeV}}^{-1}$ at the time of recombination).  Hence, on Hubble time scales the residual ionization must be taken into account, and it also allows for the inverse cascade of the magnetic field to proceed.

While our paper was in review,  an article \cite{Rama} appeared argueing that the residual ionization cannot be neglected. We do not agree with the analysis for two reasons. Firstly, on the time scale of the resonance,  plasma effects should be negligible as argued above. Secondly,  in \cite{Rama} plasma equations which hold for DC currents were used, while in our case the induced currents are AC currents.  



\end{document}